\documentclass[10pt]{article}

\usepackage{charter} 
\usepackage{fullpage}
\usepackage{titlesec}
\usepackage{amsmath,graphicx}
\usepackage{amssymb}
\usepackage{cite}
\usepackage{amsthm}
\usepackage{float}
\usepackage{color,subcaption}
\usepackage{bm}
\usepackage{bbm}
\usepackage{algorithm}
\usepackage{algpseudocode}
\usepackage{enumitem}

\usepackage{hyperref}       
\usepackage{url}            
\usepackage{booktabs}       
\usepackage{amsmath}		
\usepackage{amsthm}			
\usepackage{amsfonts}       
\usepackage{amssymb}        
\usepackage{microtype}      
\usepackage{bm}				
\usepackage{cite}			
\usepackage{graphicx}		
\usepackage{mathtools}		
\usepackage{algorithm}
\usepackage{algpseudocode}
\usepackage{multirow}       
\usepackage{tabularx}	    
\usepackage{hhline}
\usepackage{arydshln}		

\usepackage{xcolor}

\usepackage{enumitem}		
\titleformat*{\section}{\large\bfseries}
\titleformat*{\subsection}{\normalsize\bfseries}
\titleformat*{\subsubsection}{\small\bfseries}

\newcolumntype{L}[1]{>{\raggedright\arraybackslash}p{#1}}
\newcolumntype{C}[1]{>{\centering\arraybackslash}p{#1}}
\newcolumntype{R}[1]{>{\raggedleft\arraybackslash}p{#1}}

\theoremstyle{plain} 

\newtheorem{theorem}{Theorem}

\newtheorem{assumption}{Assumption}

\def\defn{\,\coloneqq\,}
\def\argmin{\mathop{\mathsf{arg\,min}}} 

\def\lim{\mathop{\mathsf{lim}}} 
 
\def\max{\mathop{\mathsf{max}}}

\def\zer{\mathsf{zer}}
\def\fix{\mathsf{fix}}

\def\DnCNNast{{\text{DnCNN}^\ast}}

\def\ebm{{\bm{e}}}

\def\sbm{{\bm{s}}}
\def\xbm{{\bm{x}}}

\def\ybm{{\bm{y}}}

\def\zerobm{\bm{0}}
\def\Abm{{\bm{A}}}
\def\Dbm{{\bm{D}}}

\def\xbmast{{\bm{x}^\ast}}

\def\xbmhat{{\widehat{\bm{x}}}}

\def\Tsf{{\mathsf{T}}}
\def\Ssf{{\mathsf{S}}}
\def\Dsf{{\mathsf{D}}}

\def\Hsf{{\mathsf{H}}}

\def\Gsf{{\mathsf{G}}}
\def\Isf{{\mathsf{I}}}

\def\Hsf{{\mathsf{H}}}

\def\R{\mathbb{R}}

\def\N{\mathbb{N}}

\begin{document}

\title{ Infusing Learned Priors into Model-Based Multispectral Imaging}

{\normalsize\author{Jiaming Liu\(^1\), Yu~Sun\(^{2}\) Ulugbek~S.~Kamilov\(^{1,2}\) \thanks{This work was supported in part by NSF grant CCF-1813910.}\\
\emph{\small \(^1 \) Department of Electrical and Systems Engineering,~Washington University in St.~Louis, MO 63130, USA.}\\
\emph{\small \(^2 \) Department of Computer Science and Engineering,~Washington University in St.~Louis, MO 63130, USA.}\\
\small\emph{email}: \texttt{\{jiaming.liu,sun.yu,kamilov\}@wustl.edu}
}}

\date{}
\maketitle 

\begin{abstract}
We introduce a new algorithm for regularized reconstruction of multispectral (MS) images from noisy linear measurements. Unlike traditional approaches, the proposed algorithm regularizes the recovery problem by using a prior specified \emph{only} through a learned denoising function. More specifically, we propose a new accelerated gradient method (AGM) variant of regularization by denoising (RED) for model-based MS image reconstruction. The key ingredient of our approach is the three-dimensional (3D) deep neural net (DNN) denoiser that can fully leverage spationspectral correlations within MS images. Our results suggest the generalizability of our MS-RED algorithm, where a single trained DNN can be used to solve several different MS imaging problems.\end{abstract}


\section{Introduction}
\label{Sec:Intro}

Multispectral (MS) imaging systems acquire the response from an object or a scene over a wide range of frequency bands, including \emph{optical}, \emph{infra-red}, and \emph{short-wave infra-red}~\cite{Willett.etal2014, Arce.etal2014}. Multi-band spectra provide rich information for detecting and classifying materials, especially those that have similar visible colors. Additionally, the higher transmission property of infra-red bands, when compared to optical bands, makes the MS imaging beneficial in imaging through haze or fog. As a result, multispectral or even hyperspectral imaging techniques have gained popularity in a number of important applications in astronomy, agriculture, and medical imaging~\cite{Solomon.Rock1985, Chen.etal2002, Hege.etal2004, Lu.Fei2014}.

While MS imaging has a great potential, acquiring and processing corresponding data is challenging due to various hardware limitations and the high-dimensional nature of typical datasets. In particular, the resolution and signal-to-noise ratio (SNR) of MS images is often constrained by limitations on the size, weight, and power of sensors used for data acquisition. Image reconstruction methods have been developed for mitigating such hardware limitations by using advanced priors on the unknown MS image. Traditional priors used in MS image reconstruction include sparsity-based regularizers, low-rank models, and dictionary-learning-based priors~\cite{Othman.Qian2006, Charles.etal2011, Lanaras.etal2015, Kanatsoulis.etal2018, Wang.etal2018}. Recently, however, the focus in the field has been shifting towards \emph{deep learning} techniques that are based on learning the direct mapping from the measurements to the recovered MS image~\cite{Hu.etal2017, Xiong.etal2017, Wen.etal2018, Liu.Lee2019}. However, the conceptual simplicity of end-to-end learning comes with the loss of modularity, characterized by the lack of an explicit separation between the measurement model and the prior. This limits the generalization and reuse of previously trained models.

In this paper, we develop a new MS image reconstruction method called \emph{multispectral regularization by denoising (MS-RED)} that enables the infusion of deep neural net (DNN) priors while maintaining an explicit separation between the prior and the measurement model. By building on the recently developed RED framework~\cite{Romano.etal2017, Reehorst.Schniter2019, Mataev.Elad2019, Sun.etal2019b}, our method specifies the learned prior only via a \emph{3D spatiospectral denoising} function, trained for the removal of additive white Gaussian noise (AWGN) from MS images. MS-RED naturally leverages both spatial and spectral correlations in the data, while also explicitly enforcing fidelity to the measured data. We discuss the convergence of MS-RED under a set of transparent assumptions on the data-fidelity and the denoiser, and develop its fast variant based on the accelerated gradient method (AGM). We finally demonstrate our MS-RED algortihm on MS superresolution using several denoising priors, including those based on DNNs. Our results illustrate that a single trained DNN denoiser can provide a state-of-the-art performance while solving multiple inverse problems without additional retraining.

\begin{figure}[t]
\begin{center}
\includegraphics[width=10cm]{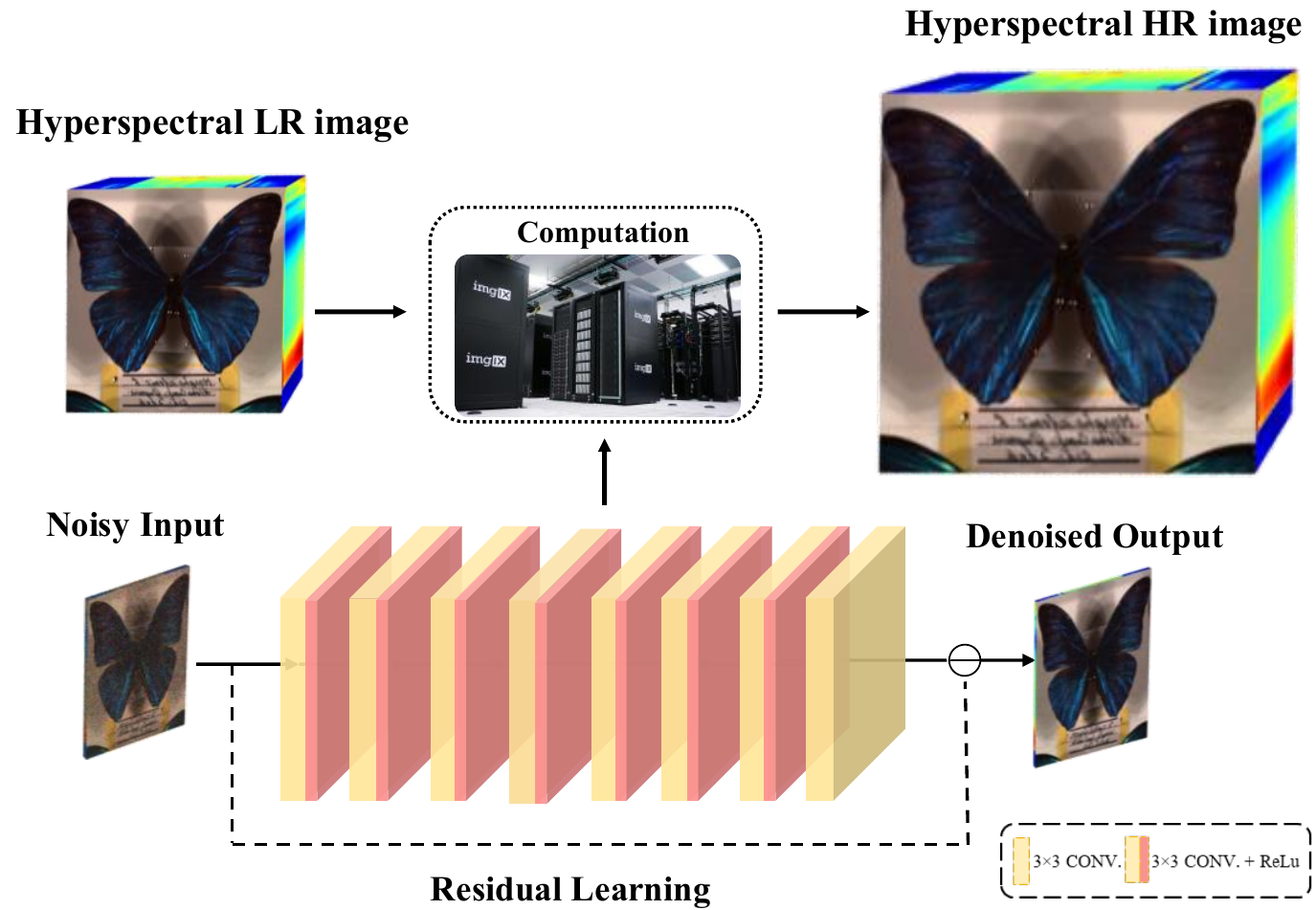}
\end{center}
\caption{We propose a new \emph{multispectral regularization by denoising (MS-RED)} algorithm for recovering a spatiospectral data-cube from its noisy measurements. Our algorithm solves this ill-posed problem by combining the measurement model with a \emph{learned} spatiospectral denoiser that maximally exploits all the available prior knowledge.}
\label{Fig:SnrCompare}
\end{figure}


\section{Proposed Method}
\subsection{Inverse Problem Formulation}

We consider the problem of acquiring a MS data cube, consisting of two spatial dimensions and one spectral dimension. The cube is represented in a vectorized form as $\xbm \in \R^n$. We additionally model the MS image acquisition process as
\begin{equation}
\label{Eq:InverseProb}
\ybm = \Abm\xbm + \ebm,
\end{equation}
where $\Abm \in \R^{m \times n}$ represents the acquisition model, $\ybm \in \R^m$ is a collection of $m$ measurements generated by our imaging system (where potentially $m \ll n$) and $\ebm \in \R^m$ is noise.

Since the recovery of $\xbm$ from $\ybm$ in~\eqref{Eq:InverseProb} is ill-posed, it is common to formulate it as a regularized optimization problem
\begin{equation}
\label{Eq:Optimization}
\xbmhat = \argmin_{\xbm \in \R^n} \left\{f(\xbm)\right\}\text{ with }f(\xbm) = g(\xbm) + h(\xbm),
\end{equation}
where $g$ is the data-fidelity term that accounts for the measurement model and $h$ is the regularizer that infuses the prior information. For example, when the noise in~\eqref{Eq:InverseProb} is AWGN, one often sets $g(\xbm) = \frac{1}{2}\|\ybm-\Abm\xbm\|_2^2$. On the other hand, sparsity-based regularization is obtained by setting $h(\xbm) = \tau \|\Dbm\xbm\|_1$, where $\Dbm$ is a some spatiospectral transform and $\tau > 0$ is the regularization parameter~\cite{Othman.Qian2006, Wang.etal2018}.

\begin{figure*}
\begin{minipage}[t]{.5\textwidth}
\begin{algorithm}[H]
\caption{$\mathsf{GM}/\mathsf{AGM}$}\label{alg:agm}
\begin{algorithmic}[1]
\State \textbf{input: } $\xbm^0 = \sbm^0$, $\gamma > 0$, and $\{q_k\}_{k \in \N}$
\For{$k = 1, 2, \dots$}
\State $\xbm^k \leftarrow \sbm^{k-1}-\gamma \nabla f(\sbm^{k-1})$\\
\quad\quad\quad where  $\nabla f(\xbm) \defn \nabla g(\xbm) + \nabla h(\xbm)$
\State $\sbm^k \leftarrow \xbm^k + ((q_{k-1}-1)/q_k)(\xbm^k-\xbm^{k-1}) $
\EndFor\label{euclidendwhile}
\end{algorithmic}
\end{algorithm}%
\end{minipage}
\hspace{0.25em}
\begin{minipage}[t]{.5\textwidth}
\begin{algorithm}[H]
\caption{$\mathsf{MS}$-$\mathsf{RED}$}\label{alg:redagm}
\begin{algorithmic}[1]
\State \textbf{input: } $\xbm^0 = \sbm^0$, $\gamma, \tau > 0$, $\{q_k\}_{k \in \N}$, and $\Dsf(\cdot)$
\For{$k = 1, 2, \dots$}
\State $\xbm^k \leftarrow \sbm^{k-1}-\gamma \Gsf(\sbm^{k-1})$\\
\quad\quad\quad where  $\Gsf(\xbm) \defn \nabla g(\xbm) + \tau(\xbm-\Dsf(\xbm))$
\State $\sbm^k \leftarrow \xbm^k + ((q_{k-1}-1)/q_k)(\xbm^k-\xbm^{k-1}) $
\EndFor\label{euclidendwhile}
\end{algorithmic}
\end{algorithm}%
\end{minipage}
\end{figure*}

\emph{Accelerated gradient method (AGM)}, summarized in Algorithm~\ref{alg:agm}, is a well-studied iterative method, effective in large-scale setting, for solving~\eqref{Eq:Optimization} when $f$ is a differentiable function~\cite{Nesterov2004, Beck.Teboulle2009b}. It can be further extended to circumvent the need to differentiate $h$ by using proximal operators~\cite{Bioucas-Dias.Figueiredo2007, Beck.Teboulle2009, Kamilov2017}.  Note that the switch between the standard \emph{gradient method (GM)} and AGM in Algorithm~\ref{alg:agm} is controlled via the sequence $\{q_k\}$. When $q_k \leftarrow 1$ for all $k$, we obtain GM with $O(1/t)$ convergence rate; when 
\begin{equation}
\label{Eq:Nesterov}
q_k \leftarrow \frac{1}{2}(1+\sqrt{1+4q_{k-1}^2}),
\end{equation}
the algorithms becomes AGM with $O(1/t^2)$ convergence. Here, we extend the traditional GM/AGM approach by including a learned denoising function $\Dsf(\cdot)$ that characterizes a MS image prior.

\subsection{Multispectral Regularization by Denoising}
In the spirit of AGM, we consider Algorithm~\ref{alg:redagm}, which iteratively estimates $\xbm$ by combining the usual gradient of data-fidelity term $\nabla g$ with an operator ${\Hsf(\xbm) \defn \tau (\xbm - \Dsf(\xbm))}$, where $\tau > 0$ is the regularization parameter and $\Dsf(\cdot)$ is a spactiospectral denoiser. When the algorithm converges, it converges to the vectors in the zero set of the operator $\Gsf$
\begin{equation*}
\zer(\Gsf) \defn \{\xbm \in \R^n : \Gsf(\xbm) = \nabla g(\xbm) + \tau(\xbm-\Dsf(\xbm)) = \zerobm\}.
\end{equation*}
Consider the following two sets
\begin{align*}
&\zer(\nabla g) \defn \{\xbm \in \R^n : \nabla g(\xbm) = \zerobm\}\quad\text{and}\quad\\
&\quad\quad\fix(\Dsf) \defn \{\xbm \in \R^n : \xbm = \Dsf(\xbm)\},
\end{align*}
where $\zer(\nabla g)$ is the set of all critical points of the data-fidelity and $\fix(\Dsf)$ is the set of all fixed points of the denoiser. Intuitively, the fixed points of $\Dsf$ correspond to all the vectors that are not denoised, and therefore can be interpreted as vectors that are \emph{noise-free} according to the denoiser. 

Note that if $\xbmast \in \zer(\nabla g)\cap\fix(\Dsf)$, then $\Gsf(\xbmast) = \zerobm$ and $\xbmast$ is one of the solutions of RED. Hence, any vector that is consistent with the data for a convex $g$ and noiseless according to $\Dsf$ is in the solution set. On the other hand, when $\zer(\nabla g)\cap \fix(\Dsf) = \varnothing$, then $\xbmast \in \zer(\Gsf)$ corresponds to a tradeoff between the two sets, explicitly controlled via $\tau > 0$. This explicit control in the strength of regularization is one of the key differences between RED and an alternative \emph{plug-and-play priors (PnP)} framework~\cite{Venkatakrishnan.etal2013, Sreehari.etal2016, Sun.etal2018a}.

The key benefits of our MS-RED approach is in its explicit separation of the measurement model from the prior, its ability to accommodate powerful spatiospectral denoisers (such as the ones based on DNNs) without differentiating them, and their excellent recovery performance as corroborated by our simulations in Section

\subsection{Convergence Analysis}

Under some conditions on the denoiser, it is possible to relate $\Hsf(\cdot)$ to some explicit regularization function $h$. For example, when the denoiser is locally homogeneous and has a symmetric Jacobian~\cite{Romano.etal2017, Reehorst.Schniter2019}, the operator $\Hsf(\cdot)$ corresponds to the gradient of
\begin{equation}
\label{Eq:REDReg}
h(\xbm) = \frac{\tau}{2}\xbm^\Tsf(\xbm-\Dsf(\xbm)).
\end{equation}
An alternative justification of RED for another explicit function $h$ is possible by considering minimum mean squared error (MMSE) denoisers~\cite{Bigdeli.etal2017, Reehorst.Schniter2019}. 
In this paper, we perform a fixed point analysis of MS-RED for denoisers that do \emph{not} necessarily correspond to any regularizer $h$. Our analysis complements the analysis in~\cite{Reehorst.Schniter2019} by developing an explicit worst-case convergence rate. We require three assumptions that together serve as sufficient conditions for convergence.
\begin{assumption}
\label{As:NonemptySet}
The operator $\Gsf$ is such that $\zer(\Gsf) \neq \varnothing$. There is a finite number $R_0$ such that the distance of the initial $\xbm^0 \in \R^n$ to the farthest element of $\zer(\Gsf)$ is bounded, that is
$$\max_{\xbmast \in \zer(\Gsf)} \|\xbm^0-\xbmast\|_2 \leq R_0.$$
\end{assumption}
\noindent
This assumption is related to the existence of the minimizers in the traditional optimization literature~\cite{Nesterov2004}. The next two assumptions rely on Lipschitz continuity of the data-fidelity and the denoiser.
\medskip
\begin{assumption}
\label{As:DataFitConvexity}
The function $g$ is smooth and convex. It has a Lipschitz continuous gradient with constant $L_g > 0$.
\end{assumption}

\noindent
This is a standard assumption used in gradient-based methods, including for the GM and its proximal extensions~\cite{Nesterov2004}.

\medskip
\begin{assumption}
\label{As:NonexpansiveDen}
The denoiser $\Dsf$ is Lipschitz continuous with constant $L_\Dsf = 1$.
\end{assumption}
\noindent
Since the proximal operator is nonexpansive~\cite{Parikh.Boyd2014}, it automatically satisfies this assumption. Spectral-normalization can be used to make DNNs satisfy this assumption~\cite{Sedghi.etal2019}.

\medskip
\begin{theorem}
\label{Thm:ConvThm1}
Run MS-RED for $t \geq 1$ iterations under Assumptions~\ref{As:NonemptySet}-\ref{As:NonexpansiveDen} using a fixed step-size $0 < \gamma \leq 1/(L_g+2\tau)$ and with $q_k = 1$. Then, we have
\begin{equation}
\label{Eq:BCREDConv}
\|\Gsf(\xbm^{t-1})\|_2^2 \leq \frac{(L_g+2\tau)}{\gamma t}R_0^2
\end{equation}
\end{theorem}
\noindent

A proof of the theorem is provided in Section~\ref{Sec:Proof}. Theorem~\ref{Thm:ConvThm1} establishes that the iterates of MS-RED can get arbitrarily close to $\zer(\Gsf)$ with $O(1/t)$ rate. The proof relies on the monotone operator theory~\cite{Bauschke.Combettes2017, Ryu.Boyd2016}, widely used in the context of convex optimization~\cite{Parikh.Boyd2014}.

The analysis in Theorem~\ref{Thm:ConvThm1} only provides \emph{sufficient conditions} for the convergence of MS-RED. As corroborated by our numerical studies in Section, the actual convergence of MS-RED is more general. It often holds with the Nesterov's acceleration in eq.~\eqref{Eq:Nesterov} and beyond nonexpansive denoisers. This suggests that some denoisers might be \emph{locally nonexpansive} over the set of input vectors used in testing. On the other hand, DNNs can be made \emph{globally nonexpansive} via spectral-normalization~\cite{Sedghi.etal2019}, which results in provable convergence.

\begin{figure}[t!]
\begin{center}
\includegraphics[width=8cm]{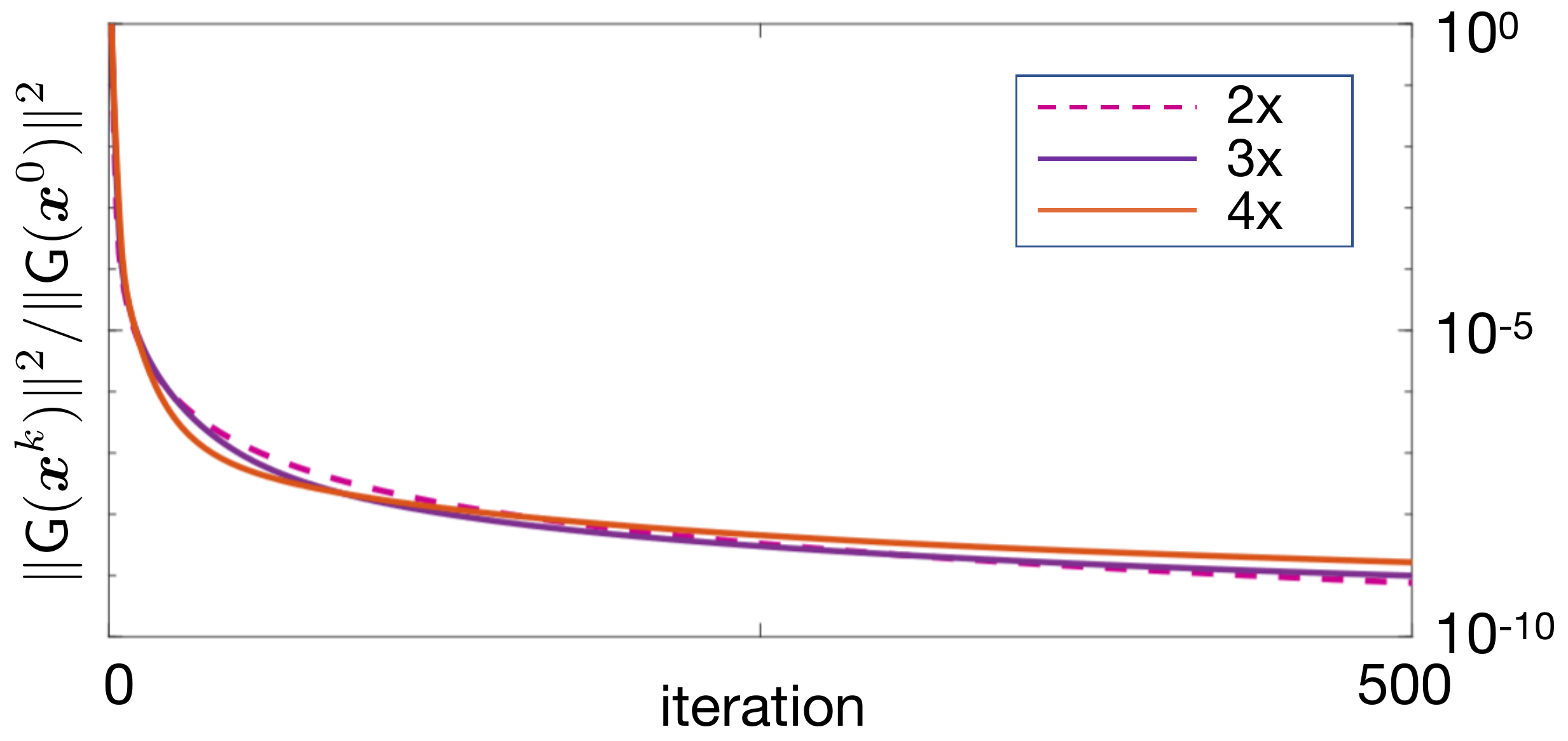}
\end{center}
\caption{Illustration of the convergence of MS-RED for three supperresolution factors, namely $\times 2$, $\times 3$, and $\times 3$. The normalized distance to $\zer(\Gsf)$ is averaged over all test images and plotted against the iteration number. The DNN used in this experiment is the residual 3D-DnCNN$^\ast$ with Lipschitz constant LC = 2. We observed the general stability of MS-RED for our denoiser across all simulations.}
\label{Fig:Convergence}
\end{figure}

\section{Numerical Experiments}
\label{Sec:Experiments}
In this section, we provide several simulations validating our theoretical analysis on \emph{multispectral image supper-resolution (MSISR)}. Generally, the low-resolution (LR) image can be modeled by a blurring and subsequent down-sampling operation on a high resolution one. We test several image denoisers, including 3DTV, BM3D, and our own DnCNN$^\ast$ on MS images from \emph{TokyoTech} database~\cite{Monno.etal2015}. Originally, those images are of size 500$\times$500$\times$31 with the wavelengths in the range of 420$\:\sim\:$720 nm at an interval of 10nm. However, for convenience, we restrict our simulations to 6 of 31 bands to build 500$\times$500$\times$6 tensors. We implement two typical image degradation settings for MSISR, namely blurring by Gaussian kernel of size $7\times7$ with $\sigma= 1.6$ and blurring by motion kernel of size $19\times19$~\cite{Liu.etal2019}, both followed by down-sampling with scale factors 2, 3 and 4. For both kernels, the measurements are corrupted with AWGN corresponding to input signal-to-noise ratio (SNR) of 40 dB.

To take advantage of the ability of DNNs to automatically learn the prior information on a specific image set, we train a \emph{residual} DNN denoiser denoted as DnCNN$^\ast$, which is a simplified version of the popular DnCNN denoiser~\cite{Zhang.etal2017}. DnCNN$^\ast$ consists of seven layers with three different blocks. The first block is a composite convolutional layer, consisting of a normal convolutional layer and a rectified linear units (ReLU) layer. It convolves the $n_1 \times n_2\times 6$ input to $n_1 \times n_2 \times 64$ features maps by using 64 filters of size $3 \times 3$. The second block is a sequence of 5 composite convolutional layers, each having 64 filters of size $3 \times 3 \times 64$. Those composite layers further processes the feature maps generated by the first part. The third block of the network, a single convolutional layer, generates the final output image with six channels by convolving the feature maps with a $3 \times 3 \times 64$ filter. Every convolution is performed with a stride $=1$, so that the intermediate feature maps share the same spatial size of the input image. Moreover, the global Lipschitz constant of the DNN is controlled to be LC = 2 via spectral-normalization ~\cite{Sedghi.etal2019}, which provides a \emph{necessary} condition for nonexpansiveness of the denoiser (\emph{i.e.}, for $L_\Dsf = 1$). We choose four MS images for testing, namely \emph{Butterfly}, \emph{Fan}, \emph{Cloth}, and \emph{Doll}. Three MS images are used for validation, and the rest twenty eight MS images are used for training. We further expand the training set by adopting patch-wise training and using standard data augmentation strategies. We train three variants of DnCNN$^\ast$ for the removal of AWGN, each using a separate noise level from $\sigma \in \{1,5,10\}$.  For each experiment, we select the denoiser providing the highest SNR performance. The BM3D denoiser was applied in a separable fashion along each MS band. The $\sigma$ parameter of BM3D was also fine-tuned for each experiment from the set $\{1,5,10\}$.

\begin{figure*}[t]
\begin{center}
\includegraphics[width=17cm]{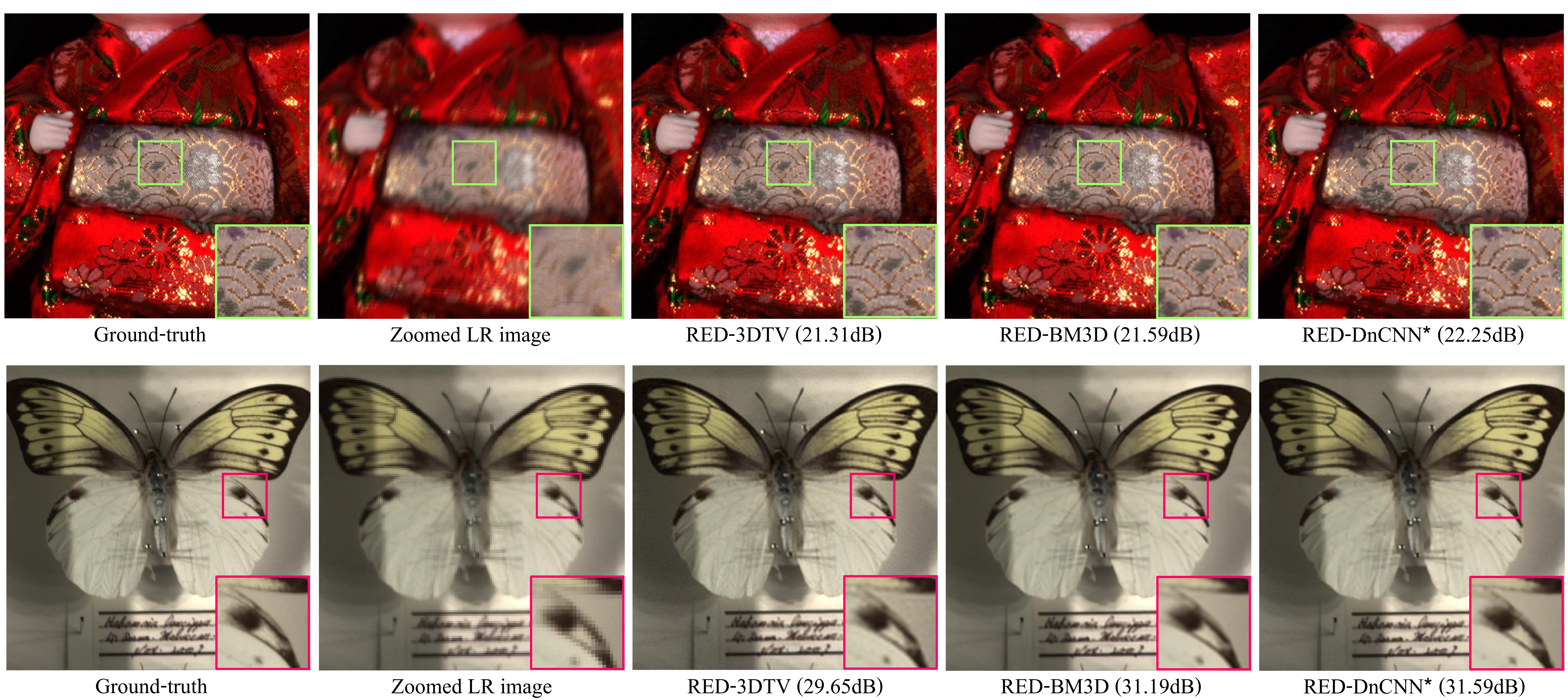}
\end{center}
\caption{Visual illustration of the reconstructed \emph{Doll} and \emph{Butterfly} images from TokyoTech dataset, obtained using MS-RED with 3DTV, BM3D, and $\DnCNNast$ denoisers. The DnCNN$^\ast$ denoiser achieves the best performance by maximally exploiting spatiospectral features in the data.}
\label{Fig:Visual}
\end{figure*}

Fig.~\ref{Fig:Convergence} illustrates the convergence of MS-RED by plotting the per-iteration normalized distance to $\zer(\Gsf)$ averaged over all the test images. Table~\ref{Tab:SNR of SR} summarizes the SNR values for different denoisers averaged over all the images. Fig.~\ref{Fig:Visual} visually illustrates the performance by forming pseudo-color images (composed of 22nd, 2nd, and 17th bands) of the results with the scale factor 2 (top) and 3 (bottom). The RED-DnCNN$^\ast$ achieves the best SNR performance while also delivering reliable convergence across all instances.

\begin{table}[t!]
\centering
\textbf{\caption{Summary of average SNR performance of different methods for multispectral image super-resolution.\label{Tab:SNR of SR}}}
\vspace{0.5em}
\scalebox{1}{
\begin{tabular}{|c|c|c|c|c|}
\hline
Kernel & Scale & 3DTV & BM3D & DnCNN*\\
\hline
\multirow{3}{5em}{\centering Gaussian\\$7\times7$}& $\times 2$ & 24.01 & 24.67 &  25.20\\ 
& $\times 3$ & 22.74 & 23.33 & 23.43\\& $\times 4$ & 20.41 & 20.94 & 21.03\\
\hline
\multirow{3}{5em}{\centering Motion blur\\$19\times19$}& $\times 2$ & 23.30 & 24.49 &  25.21\\ 
& $\times 3$ & 20.17 & 20.85 & 21.01\\& $\times 4$ & 18.86 & 19.34 & 19.43\\
\hline
\end{tabular}}
\end{table}

\section{Conclusion}
We have presented the MS-RED algorithms for reconstructing MS images from their noisy measurements. The algorithm relies on the integration of a 3D spatiospectral denoiser with the measurement model for maximally exploiting all the available information. In the case of convex data-fidelity terms and nonexpansive denoisers, MS-RED is provably convergent with the rate $O(1/t)$. However, our experiments suggest that the convergence of MS-RED extends to AGM and beyond nonexpansive denoisers. Therefore, MS-RED is applicable to a large class of MS problems and our future work will include possible extensions to different inverse problems.

\section{Proof of the Theorem}
\label{Sec:Proof}

A fixed-point convergence of averaged operators is well-known under the name of Krasnosel'skii-Mann theorem (see Section 5.2 in~\cite{Bauschke.Combettes2017}) and was recently applied to the analysis of PnP~\cite{Sun.etal2018a} and several RED algorithms in~\cite{Reehorst.Schniter2019, Sun.etal2019b}. Our analysis here extends these results to MS-RED by providing explicit worst-case convergence rates.

\medskip\noindent
In the proof below, we use the shorthand notation $L = L_g$ and $\|\cdot\| = \|\cdot\|_2$. We consider the following operators
$$\Gsf = \nabla g + \Hsf \quad\text{with}\quad \Hsf = \tau (\Isf-\Dsf).$$
and proceed in several steps by relying on fundamental results from monotone operator theory (Section 5.2 in~\cite{Bauschke.Combettes2017}).

\begin{itemize}[leftmargin=*]
\item Since $\nabla g$ is $L$-Lipschitz continuous, we know that it is $(1/L)$-cocoercive. Hence, the operator ${(\Isf-(2/L)\nabla g)}$ is nonexpansive.
\item From the definition of $\Hsf$ and the fact that $\Dsf$ is nonexpansive, we know that ${(\Isf-(1/\tau)\Hsf) = \Dsf}$ is nonexpansive.
\item We know that a convex combination of nonexpansive operators is also nonexpansive, hence we conclude that
\begin{align}
\label{Eq:Nonexpan}
&\Isf - \frac{2}{L+2\tau}\Gsf \\
\nonumber&= \left(\frac{2}{L+2\tau}\cdot \frac{L}{2}\right)\left[\Isf-\frac{2}{L}\nabla g\right] + \left(\frac{2}{L+2\tau}\cdot \frac{2\tau}{2}\right)\left[\Isf - \frac{1}{\tau}\Hsf\right],
\end{align}
is nonexpansive. This implies that that $\Gsf$ is $1/(L+2\tau)$-cocoercive.
\item Consider any $\xbmast \in \zer(\Gsf)$ and a single iteration of MS-RED $\xbm^+ = \xbm - \gamma\Gsf\xbm$. We then have
\begin{align}
\label{Eq:SingleIter}
\nonumber&\|\xbm^+-\xbmast\|^2 = \|\xbm - \xbmast - \gamma \Gsf\xbm\|^2\\
\nonumber&= \|\xbm-\xbmast\| - 2\gamma (\Gsf\xbm-\Gsf\xbmast)^\Tsf(\xbm-\xbmast) + \gamma^2\|\Gsf\xbm\|^2 \\
\nonumber&\leq \|\xbm-\xbmast\|^2 - \frac{1}{L+2\tau}\left(2\gamma-(L+2\tau)\gamma^2\right)\|\Gsf\xbm\|^2 \\
&\leq \|\xbm-\xbmast\|^2 - \frac{\gamma}{L+2\tau}\|\Gsf \xbm\|^2,
\end{align}
where we used $\Gsf\xbmast = \zerobm$, the cocoercivity of $\Gsf$, and the fact that $0 < \gamma \leq 1/(L+2\tau)$. 
\item By rearranging the terms and averaging over $t \geq 1$ iterations, we obtain
\begin{align}
\label{Eq:PreFinal}
\frac{1}{t}\sum_{k = 1}^t \|\Gsf\xbm^{k-1}\|^2 \leq \frac{1}{t} \left[\frac{(L+2\tau)}{\gamma} R_0^2\right]
\end{align}
\item From eq.~\eqref{Eq:Nonexpan}, we know that the operation 
$$\xbm^t = \Ssf\xbm^{t-1} = \xbm^{t-1} - \gamma\Gsf\xbm^{t-1}$$
is nonexpansive for any $0 < \gamma \leq 1/(L+2\tau)$. Hence, by using this nonexpansiveness, we conclude that
\begin{align*}
\|\Gsf\xbm^t\| &= \frac{1}{\gamma}\|\xbm^{t-1}-\xbm^t\| \\
&= \frac{1}{\gamma}\|\Ssf\xbm^{t-2}-\Ssf\xbm^{t-1}\| \leq \frac{1}{\gamma}\|\xbm^{t-2}-\xbm^{t-1}\| = \|\Gsf\xbm^{t-1}\|,
\end{align*}
which together with~\eqref{Eq:PreFinal} directly leads to the result in~\eqref{Eq:BCREDConv}.
\end{itemize}


\end{document}